\documentclass[
 aip,
 jcp,
 amsmath,amssymb,
 reprint
]{revtex4-2}
\usepackage{adjustbox}
\usepackage{caption}
\usepackage{dcolumn}
\usepackage{bm}
\usepackage[T1]{fontenc}
\usepackage[utf8x]{inputenc}
\usepackage[version=4]{mhchem} 
\usepackage{multirow}
\usepackage{physics}
\usepackage{soul}
\usepackage{subfig}
\usepackage{tikz}
\usepackage{threeparttable}
\usepackage[normalem]{ulem}
\usepackage{xcolor}

\newcommand{\Lower}[1]{\smash{\lower 1.5ex \hbox{#1}}}

\usepackage[colorinlistoftodos]{todonotes}

\begin{document}


\title{Efficient and reliable modeling of large $\pi$-electron systems with the Pariser--Parr--Pople Hamiltonian and pCCD-based methods 
}

\author{Zahra Karimi}
\affiliation{Institute of Physics, Faculty of Physics, Astronomy, and Informatics, Nicolaus Copernicus University in Toruń, Grudziądzka 5, 87-100 Toru\'{n}, Poland.}

\author{Somayeh Ahmadkhani}%
\affiliation{Institute of Physics, Faculty of Physics, Astronomy, and Informatics, Nicolaus Copernicus University in Toruń, Grudziądzka 5, 87-100 Toru\'{n}, Poland.}
\affiliation{Present address: Department of Mathematics and Computer Science, Freie Universität in Berlin, Germany}

\author{Katharina Boguslawski}
\affiliation{Institute of Physics, Faculty of Physics, Astronomy, and Informatics, Nicolaus Copernicus University in Toruń, Grudziądzka 5, 87-100 Toru\'{n}, Poland.}
\email{k.boguslawski@umk.pl}
\author{Paweł Tecmer}
\email{ptecmer@umk.pl}
\affiliation{Institute of Physics, Faculty of Physics, Astronomy, and Informatics, Nicolaus Copernicus University in Toruń, Grudziądzka 5, 87-100 Toru\'{n}, Poland.}
 
\date{\today}

\begin{abstract}
Model Hamiltonians offer a cost-effective way to capture the key physics of large $\pi$-conjugated systems.
In this work, we combine the Pariser--Parr--Pople (PPP) model Hamiltonian with pair Coupled Cluster Doubles (pCCD)-based methods to study the excited-state electronic structures of polycyclic aromatic hydrocarbons (PAHs).
The model Hamiltonian is implemented in the open-source PyBEST software package, which provides numerous pCCD-type models that have been shown to perform well for the electronic structure properties of large organic molecules when combined with quantum-chemical Hamiltonians and all-electron basis sets.
Within the PPP model, we probe canonical Hartree--Fock and natural pCCD-optimized orbitals in predicting excited-state properties using the linear response formalism on top of pCCD. 
Their performance is compared with configuration-interaction-based methods and the conventional EOM-CCSD approach.
Our study demonstrates that pCCD/PPP-based approaches are an efficient, cost-effective, and accurate framework to compute excited-state properties in large $\pi$-conjugated organic systems, such as extended PAHs or other systems relevant to organic electronics.
\end{abstract}

\maketitle

%
%
\section{Introduction}\label{sec:introduction}
The reliable modeling of the electronic structures and properties of large $\pi$-electron systems, such as organic semiconductors and other conjugated molecules, remains a real challenge for present-day computational physics, chemistry, and materials science.~\cite{dft-organic-electronics-range-separation, cui2020recent, risko-cr-2023}
Such systems typically exhibit a non-negligible amount of strong (static) electron correlation effects~\cite{bartlett_1994,entanglement_letter} due to their near-degenerate $\pi$-electron networks.
Furthermore, their considerable size limits the application of standard electronic structure methods, especially those suitable for strongly-correlated systems.~\cite{ppp-dmrg-polyene-chains-jcp-1998, hachmann_2007, gergo-model-hamiltonian-dmrg-pccp-2016, motta-zhang-2018, pccd-delaram-rsc-adv-2023, pccd-perspective-jpcl-2023, risko-cr-2023}
Examples for state-of-the-art computational approaches to model organic $\pi$-electron networks are truncated configuration interaction (CI) methods, such as configuration interaction singles (CIS) and configuration interaction singles and doubles (CISD), and the multi-configurational self-consistent field (MC-SCF) methods.~\cite{roos_casscf,knowles_1985,szalay2012}
These approaches can capture important features of excited states and potential energy surfaces, but still suffer from the rapid (combinatorial) growth in the number of configurations with system size and from size-consistency problems when the orbitals are not optimized in the correlation treatment. 
The orbital optimization of an approximate/truncated wave function provides a remedy, improving the accuracy in energy and property calculations.
An alternative to reduce computational cost is to move to a density matrix renormalization group (DMRG) treatment, ideally including some post-DMRG correction for dynamical correlation.~\cite{white, white2, white-qc, marti2010b, barcza_11, ors_springer, ors_ijqc, dmrg-review-garnet-chan-arpc-2011, wouters-review-dmrg-epjd-2014, gergo-model-hamiltonian-dmrg-pccp-2016, dmrg-book-1, dmrg-geom-opt, cuo_dmrg} 
On the other hand, single-reference quantum chemical methods, such as density functional theory (DFT)~\cite{parr-1989} or coupled-cluster (CC) theory~\cite{cizek_jcp_1966,cizek-diagrams-adv-chem-phys-1969, cizek_paldus_1971, paldus_cizek_shavitt_1972,bartlett_2007} are typically computationally more attractive when modeling large-scale systems.
However, they often struggle with quasi-degenerate states and cannot capture the essential physics of $\pi$-extended systems.~\cite{dft-tddft-failure-extended-systems-jcp-2002, tortorella-dft-failure-osc-jpcm-2016}

One strategy to address these challenges is switching to simplified yet physically meaningful model Hamiltonians that reduce computational cost while retaining the essential physics of the system.
The H\"uckel model \cite{huckel-1931, huckel-1932} provides a tight-binding description of $\pi$-electron systems based solely on one-electron terms.
The Hubbard model,~\cite{hubbard-1963} introduced to describe electron interactions in narrow energy bands, extends this by including an on-site Coulomb repulsion term, thereby accounting for short-range electron correlation in a minimal way.
The Pariser--Parr--Pople (PPP) model, developed by Pariser and Parr~\cite{pariser-1953a, pariser-1953b} and Pople,~\cite{pople-1953} extends the Hubbard model further by explicitly incorporating electron-electron interactions via a semi-empirical parameterization.
The theoretical foundations and improved parameterizations for conjugated systems have been established over the years.~\cite{jug-1990,ppp-parameters-barford-cpl-1998, ccd-multiple-solutions-tca-2003, sony2007polyacenes, chiappe-2015, jorner-ppp-invert-gap-2024}
Adamowicz and co-workers~\cite{Adamowicz-2014} proposed a localized semi-empirical coupled cluster framework that utilizes ethylene-like $\pi$-molecular orbitals, enabling efficient computation of (hyper)polarizabilities in nonalternant hydrocarbons.

To further enhance the efficiency of electronic structure calculations with the PPP model, we can combine them with 
geminal-based methods.~\cite{surjan1999geminals, surjan2012, gvb-pt2-xu-jcp-2013, gvb-cc-jctc-2014, johnson2017strategies, johnson2020richardson, pawel-pccp-geminal-review-2022, cc-pairing-lehtola-gordon-mp-2025}
One promising geminal model is the Antisymmetric Product of 1-reference Orbital Geminals (AP1roG), also known as pair Coupled Cluster Doubles (pCCD).~\cite{limacher-ap1rog-jctc-2013, pccd-tamar-jcp-2014, oo-ap1rog}
When combined with an efficient orbital optimization protocol~\cite{oo-ap1rog, ps2-ap1rog, piotrus_mol-phys, ap1rog-non-variational-orbital-optimizarion-jctc, pccd-tamar-jcp-2014} and dynamic energy corrections,~\cite{frozen-pccd, piotrus_pt2, garza-pccp-wdv, ap1rog-lcc, pccd-ptx, ola-tcc, seniority-zero-aqc-2024} pCCD-based approaches emerge as versatile and effective tools for modeling large-scale molecules, including organic building blocks and complexes containing transition metals, lanthanides, and actinides.~\cite{pawel_jpca_2014, pawel-pccp-2015, post-pccd-entanglement, pccd-ptx, pawel-yb2, pccd-ee-f0-actinides, ola-qit-actinides-pccp-2022, pccd-ci, ola-tcc, pccd-static-embedding, pccd-perspective-jpcl-2023, pccd-mocco-galynska-pccp-2024, ea-eom-pccd-jpca-2024, dip-fpccsd-jctc-2025, ea-eom-pccd-frozen-pair-jctc-2025, eom-fpccsd-faraday-disccus-2026}
Recent quantum chemical calculations on organic electronic molecules~\cite{pccd-delaram-rsc-adv-2023, pccd-perspective-jpcl-2023, ea-eom-pccd-jpca-2024, pccd-charge-transfer-jctc-2025, pccd-ip-ea-mod-koopmans-jcp-2025, dip-fpccsd-jctc-2025, pccd-doping-organic-dyes-jpca-2025, ip-ekt-delaram-jctc-2026, pccd-in-pccd-jcp-2026} show remarkably good predictive power at lower computational cost compared to standard electronic structure methods.
Most importantly, orbital-optimized pCCD-based methods improve upon the canonical Hartree--Fock orbitals in quasi-degenerate systems such as for the one-dimensional Hubbard model,~\cite{oo-ap1rog} long polymer chains, or $\pi$-extended systems.~\cite{pccd-delaram-rsc-adv-2023, ea-eom-pccd-jpca-2024,pccd-charge-transfer-jctc-2025}
To that end, in this work, we combine the PPP model Hamiltonian with pCCD-based methods.
Our application focuses on modeling the electronic structures and UV-Vis spectra of selected excited states in polycyclic aromatic hydrocarbons (PAHs), for which reliable theoretical and experimental data are available.
\section{Theory}\label{sec:theory}
\subsection{The PPP Model Hamiltonian}\label{sec:model-comput-det}
The PPP model Hamiltonian, commonly used to describe the electronic structure of conjugated $\pi$-electron systems composed of carbon atoms only, is given by
\begin{align}\label{eq:ppp}
    \hat{H}_{\text{PPP}} &=
        \sum_{\langle i,j \rangle, \sigma} t_{ij} \left( \hat{c}_{i\sigma}^\dagger \hat{c}_{j\sigma}
        + \hat{c}_{j\sigma}^\dagger \hat{c}_{i\sigma} \right)
        + U\sum_i \hat{n}_{i\uparrow} \hat{n}_{i\downarrow} \nonumber \\
        &+ \sum_{i < j} V_{ij} \left(\hat{n}_i - 1\right)\left(\hat{n}_j - 1\right),
\end{align}
where the terms represent distinct physical interactions within the $\pi$-electron system.
The first term in Eq.~\eqref{eq:ppp} describes the kinetic energy with nearest neighbor hopping parameter $t_{ij}$.
The notation ${\langle i,j \rangle}$ in the summation indicates the nearest-neighbor approximation. 
The $\hat{c}_{i\sigma}^\dagger$ and $\hat{c}_{j\sigma}$ are electron creation and annihilation operators at sites $i$ and $j$, respectively, with spins $\sigma$ ($\uparrow$ or $\downarrow$).
The second term  models the electron–electron repulsion between two $\pi$-orbitals located on sites $i$ and $j$ via a distance-dependent Coulomb integral (cf. Eq.~\eqref{eq:mod-ohno}) with $\hat{n}_i=\hat{n}_{i\uparrow}+\hat{n}_{i\downarrow}$, where $\hat{n}_{i\sigma} = \hat{c}_{i\sigma}^\dagger \hat{c}_{i\sigma}$ is the particle-number operator (again $\sigma \in \{\uparrow, \downarrow\}$), denoting the number of electrons at sites $i$.
The parameter $V_{ij}$ represents the elements of the interaction matrix and the diagonal elements $V_{ii}=U$ describe on-site interactions present in the third term at site $i$ with opposite spins.

A commonly used Ohno~\cite{ohno1964some} parametrization for the parameter $V_{ij}$ reads
\begin{equation}
V_{ij} = \frac{U}{\kappa_{ij} \sqrt{1 + 0.6117 \, R_{ij}^2 }},
\label{eq:mod-ohno}
\end{equation}
where $\kappa_{ij}$ encodes dielectric screening, incorporating effects such as electronic delocalization, polarization, and environmental screening.
We note that the on-site and intersite screening factors are distinguished explicitly, with $\kappa_{ii} = 1$ and $\kappa_{ij} \neq 1$ for $i \ne j$ for the on-site Coulomb interaction, respectively.

In Eq.~(\ref{eq:mod-ohno}), we adopt the zero-differential-overlap (ZDO) approximation simulating an orthogonal $p_{z}$ atomic-orbital basis.~\cite{pople-1953, pariser-1953a}
This simplifies the two-electron interactions by retaining only Coulomb integrals and neglecting off-diagonal two-electron terms, but preserving the essential physics,
\begin{equation}
S_{ab} = \int \chi_a(\mathbf{r}) \chi_b(\mathbf{r})\, d\mathbf{r} \approx 0 \quad \text{for } a \ne b.
\label{eq:overlap}
\end{equation}

In the present PPP parametrization, the on-site interaction $U$ and the long-range Coulomb term $V_{ij}$ are treated independently, which allows us to reproduce experimental excitation energies while retaining the flexibility of the model.
Such a parameterization has been successfully used in various electronic structure methods, including the multi-reference singles and doubles configuration interaction (MRSDCI) methodology ~\cite{bhattacharyya-2020, sony2007polyacenes, chakraborty-2013} as well as coupled cluster approaches.~\cite{Alavie2025ccsdt}


\subsection{pCCD-based Methods}\label{sec:comput-details}
Coupled cluster theory~\cite{cizek_jcp_1966, cizek-diagrams-adv-chem-phys-1969, cizek_paldus_1971} with single and double excitations (CCSD) is widely employed to describe dynamic (weak) electron correlation.~\cite{bartlett_1994, bartlett_2007}
However, it often fails in systems with strong (static, non-dynamic) electron correlation.~\cite{entanglement-jctc-2013, oo-ap1rog,post-pccd-entanglement}
Such multi-reference problems can be modeled within a multi-reference formulation of CC theory.~\cite{adamowicz-mrcc,bogus_mrcc,monika_mrcc,Koehn2013}
A cheaper alternative to address the limitations of single-reference CC theory is to simplify the CCSD ansatz even further.
One such simplified variant is the pair Coupled Cluster Doubles (pCCD)~\cite{limacher-ap1rog-jctc-2013} method that restricts the cluster operator to electron-pair excitations (with opposite spins) only.~\cite{pawel-pccp-geminal-review-2022}
This restriction significantly reduces computational cost while enabling a reliable treatment of static electron correlation,~\cite{oo-ap1rog, pawel_jpca_2014, pawel-pccp-2015,pccd-prb-2016} including large organic molecules,~\cite{ea-eom-pccd-frozen-pair-jctc-2025, dip-fpccsd-jctc-2025, pccd-ip-ea-mod-koopmans-jcp-2025} and $\pi$-extended systems.~\cite{pccd-delaram-rsc-adv-2023, ea-eom-pccd-jpca-2024}

The pCCD wave function~\cite{limacher-ap1rog-jctc-2013, pccd-tamar-jcp-2014, oo-ap1rog, pawel-pccp-geminal-review-2022} is defined as
\begin{equation}\label{eq:pccd}
    \ket{\Psi_{\rm {pCCD}}} = e^{\hat{T}_{\rm p}}|\Phi_0\rangle,
\end{equation}
where $|\Phi_0\rangle$ is some reference Slater determinant and $\hat{T}_{\rm p}$ is the cluster operator that includes only seniority-zero excitations (electron pairs),
\begin{equation}\label{eq:tp}
    \hat{T}_{\rm p} = \sum_i^{\rm occ} \sum_a^{\rm virt} t_{i_\uparrow i_\downarrow}^{a_\uparrow a_\downarrow} \hat{c}_{a_\uparrow}^\dagger \hat{c}_{a_\downarrow}^\dagger \hat{c}_{i_\downarrow} \hat{c}_{i_\uparrow}.
\end{equation}
In the above equation, the sums run over all occupied $i$ (virtual $a$) orbital (sites) of the reference determinant $\ket{\Phi_0}$.
The corresponding pCCD energy is given by (for any Hamiltonian $\hat H$)
\begin{equation}
    E = \langle\Phi_0|e^{-\hat{T}_{\rm p}}\hat{{H}}e^{\hat{T}_{\rm p}}\,|\Phi_0\rangle
\end{equation}
and the cluster amplitudes are determined from the amplitude equations by projecting against all pair-excited determinants $\bra{\Phi_{i_\uparrow i_\downarrow}^{a_\uparrow a_\downarrow}} = \bra{\Phi_0} \hat{c}_{i_\uparrow}^\dagger \hat{c}_{i_\downarrow}^\dagger \hat{c}_{a_\downarrow} \hat{c}_{a_\uparrow}$,
\begin{equation}
    \langle\Phi_{i_\uparrow i_\downarrow}^{a_\uparrow a_\downarrow}|e^{-\hat{T}_{\rm p}}\hat{{H}}e^{\hat{T}_{\rm p}}\,|\Phi_0\rangle = 0.
\end{equation}
pCCD is an efficient reparameterization of a full configuration interaction (FCI) solution in the seniority-zero sector, provided the orbitals are optimized to define the pairing structure.~\cite{oo-ap1rog, ps2-ap1rog, ap1rog-non-variational-orbital-optimizarion-jctc, piotrus_mol-phys, pccd-tamar-jcp-2014}
This orbital optimization is crucial to maximize accuracy when modeling strong electron correlation and to recover size-consistency of potential energy surfaces.~\cite{oo-ap1rog, pawel-pccp-2015, pccd-prb-2016, pccd-ptx, piotrus_pt2}
Besides, the natural (localized) pCCD orbitals serve their purpose well in the charge transfer analysis of excited states.~\cite{pccd-delaram-rsc-adv-2023,pccd-charge-transfer-jctc-2025, eom-ccsd-ct-jpcl-2026, ea-eom-pccd-frozen-pair-jctc-2025}

Unlike Hartree--Fock (HF) orbitals, which are typically delocalized over the entire $\pi$-system, the variationally optimized pCCD orbitals often become localized, mimicking the behavior of individual electron pairs (or Lewis pairs) in model and realistic molecular systems.~\cite{oo-ap1rog, ps2-ap1rog, pccd-prb-2016, pccd-delaram-rsc-adv-2023}
During orbital optimization, the orbitals are rotated to minimize the total energy of the correlated pCCD wave function --- leading to an orbital basis that reflects the underlying pairing structure of the system.
Consequently, the effective one-electron parameters (for instance, hopping integrals and orbital energies in model Hamiltonians) will vary with the chosen orbital basis (canonical HF vs. natural pCCD).
The orbital basis dependence is an intrinsic feature of any approximate wave function model, including the pCCD ansatz.~\cite{ahmadkhani2024lrpccd} 
The usefulness of orbital optimization within the pCCD wave function has been demonstrated for the 1-D Hubbard model with and without periodic boundary conditions.~\cite{oo-ap1rog, pccd-prb-2016}
To that end, the orbital-optimized pCCD approach will require a new and, ideally, unique set of parameters in the PPP model Hamiltonian. 
That differs, for instance, from the DMRG algorithm, which is invariant to unitary transformations of the orbital basis (assuming no active orbital spaces) and thus shows minimal sensitivity to the PPP model parameters.~\cite{anusooya1997-ehubbard,raghu2002-PPP, das2006-PPP-polythiophene}

Recent developments have significantly extended the applicability of pCCD-based methods to electronic excited-state calculations through the Equation-of-Motion (EOM) formalism.~\cite{eom-pccd, eom-pccd-erratum, eom-pccd-lccsd, eom-fpccsd-faraday-disccus-2026}
The linear-response formulation of pCCD (LR-pCCD) and its variants, such as LR-pCCD+S, enable efficient computations of singly and pair-excited state energies and transition properties.~\cite{ahmadkhani2024lrpccd}
We should note that the EOM and LR formalisms share the same excitation energies, but differ in the treatment of excited-state dipole moments (or other excited-state properties).~\cite{bartlett_2007}
Both methods provide a scalable, reliable approach for studying the electronic properties of the molecular building blocks of bulk heterojunction organic solar cells (OSCs), aiding the rational design of efficient organic materials.
To that end, the synergistic combination of pCCD-based methods and model Hamiltonians holds significant promise for improving scalability and predictive accuracy, thereby opening new frontiers in the investigation of large-scale quantum systems.

\subsection{Dipole Integrals in Physical Model Hamiltonians}
In semi-empirical approaches, such as the PPP model with Ohno parametrization, the evaluation of dipole moments does not rely on atom-centered basis functions or explicit momentum integrals. Instead, dipole integrals are constructed directly from atomic coordinates, consistent with the tight-binding formalism.

To compute the dipole moment vector $\vec{\mu}$, we need to define effective momentum integrals in the $x$, $y$, and $z$ directions as follows
\begin{equation}
(\mu_\alpha)_{ij} = S_{ij} \cdot \frac{1}{2}(r^\alpha_i + r^\alpha_j),
\quad \alpha \in \{x, y, z\},
\end{equation}
where $r^\alpha_i$ is the $\alpha$-component of the Cartesian coordinate of atom $i$, $S_{ij}$ is an overlap-like matrix (defined below), and $i,j$ denote the sites/atoms. 
This formulation incorporates both the spatial extent of $\pi$-orbitals and their mutual overlap, leading to a more accurate representation of the off-diagonal dipole moment contributions.

To account for spatial proximity and nonorthogonality between orbitals, we also define an overlap-like matrix $\textbf{S}$ using a Gaussian decay function,~\cite{daudel1960quantum}
\begin{equation}
S_{ij} = -\exp\left( -\left( \frac{|\vec{r}_i - \vec{r}_j|}{\lambda} \right)^2 \right),
\end{equation}
where $\lambda$ is a tunable decay parameter controlling the effective range of orbital overlap, which we default to 3.
In this work, we worked in the ZDO approximation and set $S_{ij} = - \delta_{ij}$.
We should stress that, in this work, the choice between ZDO and a Gaussian decay function does not affect the calculated spectra qualitatively.

The final ground state dipole moment is then computed using the electronic density matrix (or 1-particle reduced density matrix) $\rho$ as~\cite{pccd-dipole-moments-jctc-2024}
\begin{equation}
D_\alpha = \operatorname{Tr}[\mu_\alpha \cdot \rho], \quad \alpha \in \{x, y, z\}.
\end{equation}
This approach ensures a physically meaningful estimate of the dipole moment within a minimal basis model without invoking ab initio basis functions or momentum integral evaluations in an atom-centered basis set.~\cite{daudel1960quantum}

In typical conjugated systems (e.g., benzene or graphene fragments), the C--C bond length is nearly constant ($\approx$ 1.40--1.42~\AA), comprising the 2$p_z$ orbitals with a fixed spatial extent.
Thus, the overlap matrix ($S_{ij}$) between neighboring atoms will also be nearly uniform for nearest neighbors.~\cite{ohno1964some}
One can use various computational approaches, such as LR-CC or EOM-CC based methods, to obtain excited-state dipole moments.
Still, only the linear response formalism ensures that transition properties scale correctly with system size. 
To that end, we compute oscillator strengths using the linear-response method,~\cite{ahmadkhani2024lrpccd} and pCCD response density matrices to investigate the intensity of electronic transitions.~\cite{pccd-dipole-moments-jctc-2024, pccd-expectation-value-1dm-jpca-2025}

\begin{figure*}[tb]
\centering
\includegraphics[width=0.75\textwidth]{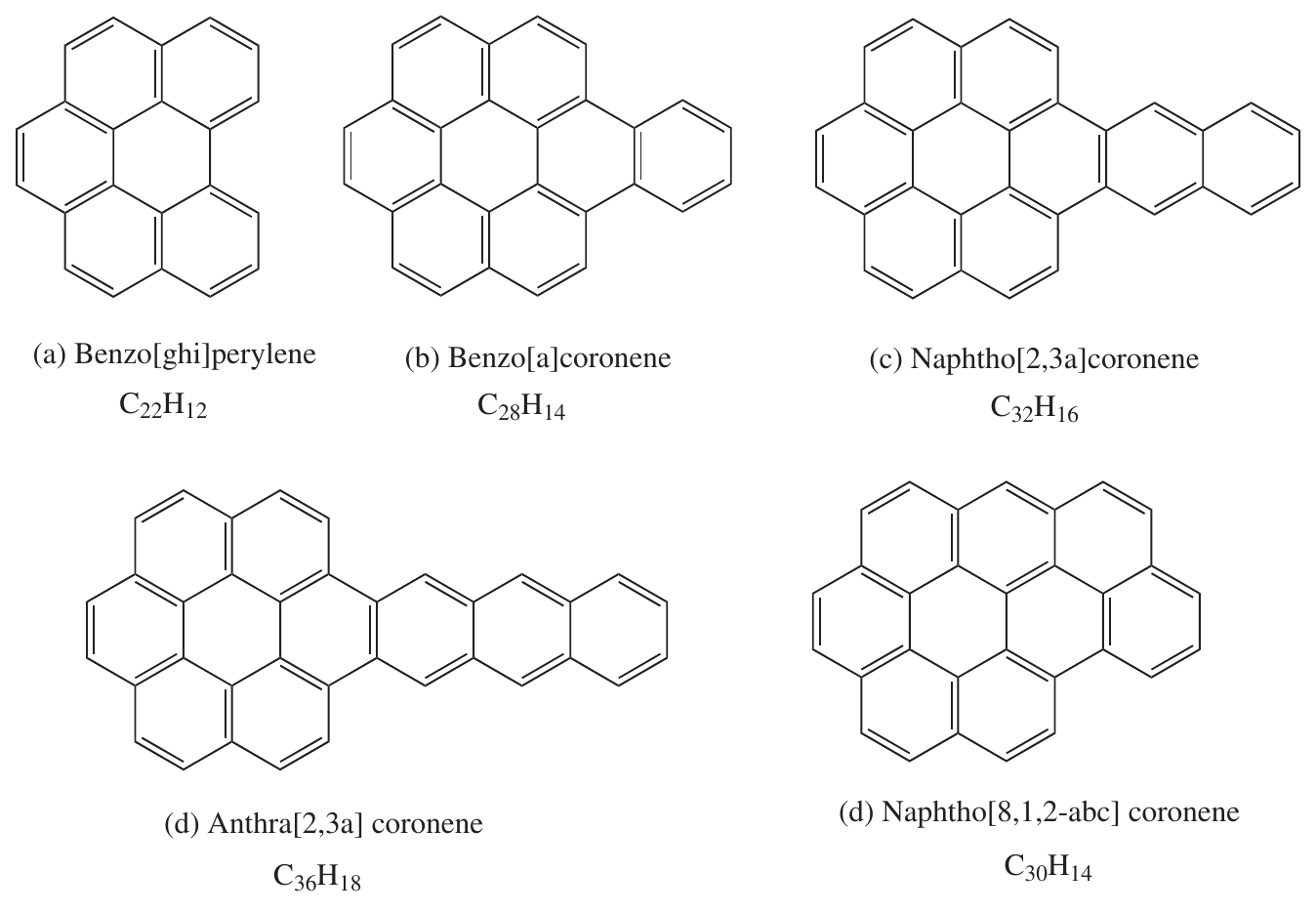}
\caption{Representative structures of polycyclic aromatic hydrocarbons (PAHs) considered in this study. 
Note that the hydrogen atoms are not explicitly shown as they are not accounted for in the PPP model. }
\label{fig:molecules}
\end{figure*}

\section{COMPUTATIONAL DETAILS}\label{sec:computational detalis}
\subsection{Geometry Optimization}\label{computational details}

In contrast to the earlier studies,~\cite{bhattacharyya-2020} where a uniform C--C bond length of 1.40~\AA{} was assumed for all structures, we employed fully optimized molecular geometries in our calculations (with the presence of hydrogen atoms).
All structures were optimized at the DFT level using the BP86~\cite{becke88} exchange--correlation functional and the cc-pVDZ basis set~\cite{basis_dunning} as implemented in ORCA.~\cite{orca-2012, orca-2018, orca-2020, orca-2022}
The obtained xyz structures are provided in the Supplementary Information.~\cite{si-info}
Harmonic vibrational frequency calculations were performed to confirm that the optimized structures correspond to true minima on the potential energy surface.
The resulting geometries were used in all subsequent pCCD-based and CCSD-based calculations using the PPP Hamiltonian recipe described in the subsequent sections.

\subsection{Refinement of Model Hamiltonian Parameters}\label{sec:pccd-comput-details}
We implemented the PPP model Hamiltonian into the open-source PyBEST software package~\cite{pybest-paper,pybest-paper-update1-cpc-2024} and used the conventional Ohno expression in eq.~\eqref{eq:mod-ohno} for the long-range interaction.
We tested our initial implementation using data from existing programs.~\cite{ppp-program-fortran90-cpc-2010, ppp-program-horton-jcp-2024}
We initially tested two commonly used parameter sets: `screened' parameters ($U = 8$~eV, $\kappa_{ij} = 2$, $\kappa_{ii}=1$, and $t = -2.4$~eV)  and the `standard' parameters ($U = 11.13$~eV, $\kappa_{ij}=\kappa_{ii} = 1$, and $t = -2.4$~eV) as widely adopted in the literature for MRCISD calculations and CASSCF orbitals.~\cite{bhattacharyya-2020, PPP-screened-parameters-prb-1997}

While both parameter sets yield reasonable results at the HF level (the computed orbital/site energies reproduce other theoretical data from Ref.~\citenum{bhattacharyya-2020} and represent a good 
approximation to the experimental optical gap of the investigated systems), the pCCD-based excitation energies computed with both RHF and pCCD-optimized orbitals show sensitivity to the choice of model parameters, motivating the investigation of an alternative parameterization.

Upon pCCD orbital optimization,~\cite{oo-ap1rog,ps2-ap1rog, ap1rog-non-variational-orbital-optimizarion-jctc} the delocalized HF orbitals become localized (symmetry broken), which affects the resulting 1- and 2-electron integrals.
Thus, we test two new parameter sets, namely ($U = 7.0$~eV, $\kappa_{ij}=\kappa_{ii} = 1$, and $t = -2.4$~eV) and ($U = 8.0$~eV, $\kappa_{ij}= 2, \kappa_{ii} = 1$, and $t = -2.4$~eV), obtained by adjusting $U$ and $\kappa$ to improve agreement with MRSDCI reference data~\cite{bhattacharyya-2020} when working with pCCD or CCSD-type methods.
In the following, we will abbreviate the set of PPP parameters using square brackets as $[t/\kappa/U]$.

\subsection{UV-Vis Spectra Simulation}
The UV-Vis absorption spectra were simulated by broadening the vertical excitation energies and oscillator strengths obtained from the LR-pCCD+S calculations.
All spectral simulations were performed using the \texttt{spec-gen.py} code of Autschbach and coworkers\cite{Moore_specgen} employing a Lorentzian broadening of $\gamma=0.15$.
In all the LR-pCCD+S calculations, the oscillator strengths were calculated from the 20 lowest-lying roots.


%
%
\section{Results}

\begin{table*}[htbp]
\centering
\caption{First vertical excitation energies (in eV) of selected PAHs computed using the PPP Hamiltonian with different screening parameterizations exploiting canonical RHF and natural pCCD-optimized orbitals.
The mean error (ME), mean absolute error (MAE), and standard deviation (SD) are given with respect to MRSDCI data.
The PPP parameters are labeled as [$t$/$\kappa$/$U$].
MRSDCI reference values are taken from Ref.~\cite{bhattacharyya-2020}. MRSDCI value for coronene 
taken from Ref.~\cite{aryanpour2014}. Experimental values are taken from Refs.~\cite{abouaf2009, bagley2013} for coronene and its derivatives, respectively. }

\vspace{0.5em}
\label{tab:eom_hf_topology}

\begin{tabular}{l|c|cc|cc|cc|cc|cc|cc|c}
\hline
Molecule & MRSDCI & \multicolumn{8}{c|}{EOM-pCCD+S} & \multicolumn{4}{c|}{EOM-CCSD} &  Exp\\
\cline{3-10} \cline{11-14}
&&
\multicolumn{4}{c|}{[-2.4/2/8.0]} &  
\multicolumn{4}{c|}{[-2.4/1/7.0]} &  
\multicolumn{2}{c|}{[-2.4/2/8.0]} &  
\multicolumn{2}{c|}{[-2.4/1/7.0]} &  
\\
\cline{3-14}
&& 
\multicolumn{2}{c|}{RHF} & \multicolumn{2}{c|}{pCCD} &
\multicolumn{2}{c|}{RHF} & \multicolumn{2}{c|}{pCCD} &
RHF & pCCD &
RHF & pCCD &
\\
&& 
EE [eV] & OS &
EE [eV] & OS &
EE [eV] & OS &
EE [eV] & OS &
EE [eV] & EE [eV] & EE [eV] & EE [eV] 
\\
\hline
$\mathrm{C}_{22}\mathrm{H}_{12}$& 3.39  &  3.18  &  0.86  &  3.52  &  0.87  &  3.10  &  0.49  &  3.14  &  0.56  &  3.30  &  3.30  &  3.01  &  3.01& --   \\
$\mathrm{C}_{24}\mathrm{H}_{12}$& 4.14 & 3.83 & 2.21 & 4.21 & 43.51 & 4.26 & 2.41 & 4.40 & 0.11 & 3.70 & 3.70 & 4.06 & 4.36 & 4.28$^{\dagger}$ \\
$\mathrm{C}_{28}\mathrm{H}_{14}$& 3.42  &  3.21  &  0.47  &  3.66  &  0.42  &  3.09  &  0.15  &  3.14  &  0.18  &  3.31  &  3.32  &  3.02  &  3.02& 3.38 \\
$\mathrm{C}_{30}\mathrm{H}_{14}$& 2.95  &  2.84  &  0.79  &  3.21  &  0.79  &  2.84  &  0.45  &  2.87  &  0.45  &  2.96  &  2.96  &  2.73  &  2.73& 3.02 \\
$\mathrm{C}_{32}\mathrm{H}_{16}$& 2.93  &  2.88  &  0.49  &  3.31  &  0.50  &  2.89  &  0.29  &  2.96  &  0.28  &  3.02  &  3.03  &  2.81  &  2.81& --   \\
$\mathrm{C}_{36}\mathrm{H}_{18}$& 2.70  &  2.54  &  0.37  &  2.97  &  0.40  &  2.61  &  0.36  &  2.69  &  0.36  &  2.75  &  2.76  &  2.51  &  2.51& --   \\
\hline
ME    && -0.15  &    &  0.26  &    & -0.17  &    & -0.12  &    & -0.01  & -0.01  & -0.26  & -0.26 \\
MAE   &&  0.15  &    &  0.26  &    &  0.17  &    &  0.13  &    &  0.07  &  0.07  &  0.26  &  0.26 \\
SD    &&  0.07  &    &  0.09  &    &  0.13  &    &  0.14  &    &  0.09  &  0.09  &  0.12  &  0.12 \\  \hline
ME (30+)  &&-0.11 &  & 0.30 &  & -0.08 &  & -0.02 &  & 0.05 & 0.06 & -0.18 & -0.18 \\
MAE (30+) && 0.11 &  & 0.30 &  &  0.08 &  &  0.04 &  & 0.05 & 0.06 &  0.18 &  0.18 \\
SD (30+)  && 0.06 &  & 0.07 &  &  0.04 &  &  0.05 &  & 0.04 & 0.04 &  0.05 &  0.05 \\
\hline

\end{tabular}
\begin{minipage}{\textwidth}
\vspace{0.5em}
\footnotesize{$^{\dagger}$Experimental values for the first allowed 
$^1E_{1u}$ transition of coronene ($\mathrm{C}_{24}\mathrm{H}_{12}$) ranging from 4.07 to 4.28~eV as 
compiled in Ref.~\cite{abouaf2009}. The value of 4.28~eV from 
electron energy loss spectroscopy in the gas phase is taken as the 
most reliable reference, as it corresponds to a direct gas-phase 
measurement. Other experimental values ranging from 4.07 to 4.25~eV 
have been reported in Refs.~\cite{Ohno1972,Nijegorodov2001,
Khakoo1990,Joblin1992,Schmidt1977}.}
\end{minipage}
\end{table*}

\begin{figure*}[t]
\centering
\includegraphics[width=0.9\textwidth]{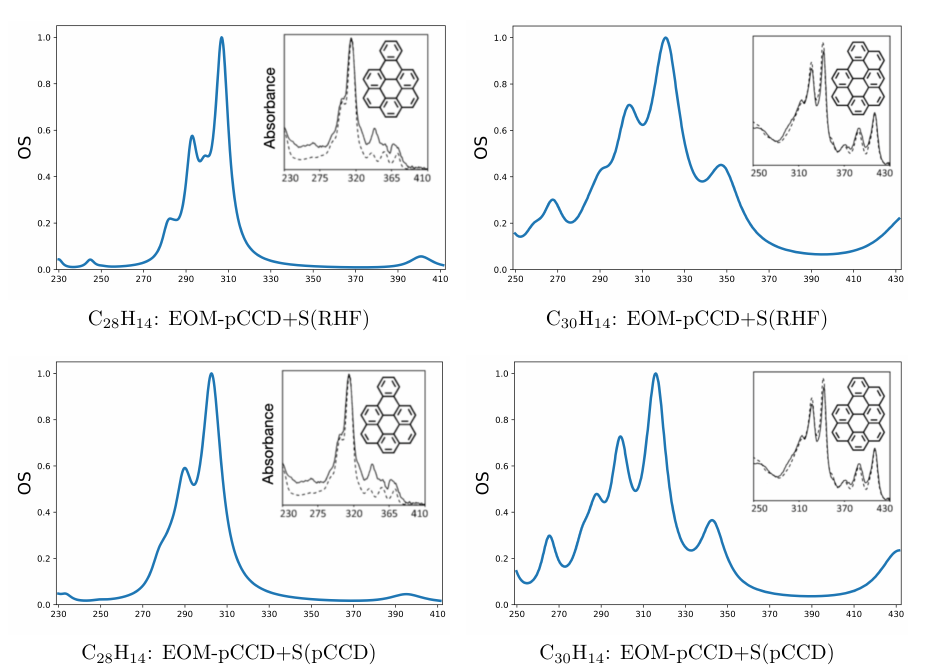}
\caption{UV absorption spectra of (left column) benzo[a]coronene and (right column) naphtho[8,1,2-abc]coronene computed using LR-pCCD+S (HF orbitals on top, pCCD-optimized orbitals on the bottom) with the PPP Hamiltonian and the parameter set  [2.4/1/7.0].
The experimental UV-Vis spectra\cite{bagley2013, abouaf_coronene_eels} is inserted for comparison.}
\label{fig:uvvis}
\end{figure*}

\subsection{Excited States with Canonical and pCCD-Optimized Orbitals}
\label{sec:results-eom}

Table~\ref{tab:eom_hf_topology} reports the first vertical excitation energies of the investigated PAHs computed using EOM-pCCD+S and EOM-CCSD with two parameter sets, [-2.4/2/8.0] and [-2.4/1/7.0], exploiting both canonical RHF and pCCD-optimized orbitals (\textit{cf}. section~\ref{sec:pccd-comput-details} for the introduced square bracket notation).

With canonical RHF orbitals, both parameter sets yield comparable accuracy for EOM-pCCD+S, with the [-2.4/2/8.0] set giving ME $= -0.15$~eV, MAE $= 0.15$~eV, and SD $= 0.07$~eV, and the [-2.4/1/7.0] set giving ME $=-0.17$~eV, MAE $= 0.17$~eV, and SD $= 0.13$~eV, both systematically underestimating the MRSDCI 
reference values.
For EOM-CCSD with RHF orbitals, the [-2.4/2/8.0] set performs significantly better (MAE $= 0.07$~eV, SD $= 0.09$~eV) than the [-2.4/1/7.0] set (MAE $= 0.26$~eV, SD $= 0.12$~eV).

Switching to pCCD-optimized orbitals has a markedly different effect depending on the parameter set.
For the [-2.4/2/8.0] set, pCCD orbitals worsen the EOM-pCCD+S results considerably, increasing the MAE from 0.15~eV to 0.26~eV with a systematic overestimation (ME $= +0.26$~eV).

We also note that the oscillator strength for coronene (\ce{C24H12}) with this parameter set and pCCD orbitals takes an anomalously large value of 43.51, which is likely a numerical artifact.
In contrast, for the [-2.4/1/7.0] set, pCCD orbitals improve the results: the MAE decreases from 0.17~eV to 0.13~eV and the ME reduces to $-0.12$~eV across all molecules.

The improvement is particularly pronounced for the larger PAHs: for molecules with 30 or more carbon atoms, EOM-pCCD+S with [-2.4/1/7.0] pCCD orbitals achieves ME $= -0.02$~eV, MAE $= 0.04$~eV, and SD $= 0.05$~eV, representing excellent agreement with MRSDCI reference data.~\cite{bhattacharyya-2020}

For EOM-CCSD, the choice of orbital basis has essentially no effect on the excitation energies: RHF and pCCD-orbital results are nearly identical across all molecules and both parameter sets.
With the [-2.4/2/8.0] set, EOM-CCSD achieves the lowest overall errors among all methods (MAE $= 0.07$~eV, SD $= 0.09$~eV), while the [-2.4/1/7.0] set systematically underestimates excitation energies by 0.26~eV on average, making it less suitable for this method.

Additional statistical analysis for systems containing more than 30 carbon atoms (denoted as 30+) is provided at the bottom of Table~\ref{tab:eom_hf_topology}.
These results clearly demonstrate that the $[-2.4/1/7.0]$ basis set combination yields the best performance for the EOM-pCCD+S method, irrespective of whether HF or pCCD reference orbitals are used.
This motivates us to employ this parameter set for modeling the UV-Vis spectra of \ce{C_{28}H_{14}} and \ce{C_{30}H_{14}} in the subsequent section.

\subsection{LR-pCCD+S and Electronic Spectra}
To in-depth analyze the electronic spectra of selected PAHs, we computed oscillator strengths using the linear-response ansatz within 
the LR-pCCD+S framework (the same excitation energies as from EOM-pCCD+S)~\cite{ahmadkhani2024lrpccd} for \ce{C_{28}H_{14}} and \ce{C_{30}H_{14}}, and compared the resulting theoretical spectra with available experimental intensity data.~\cite{PAH-exp-excite-as-1990, PAH-exp-excit-pac-2007}
Since the PPP model Hamiltonian does not involve an explicit basis set, the first step in obtaining oscillator strengths was to compute site-dependent momentum integrals (the theoretical details of this approach are outlined in Section~\ref{sec:theory}).
The resulting UV--Vis spectra are shown in Figure~\ref{fig:uvvis}.
It should be noted that the experimental UV--Vis spectra for benzo[a]coronene and naphtho[8,1,2-abc]coronene were measured in the liquid phase using HPLC coupled with UV--Vis spectroscopy and mass spectrometry.~\cite{bagley2013}
Since our calculations are performed on isolated molecules in the gas phase, solvent effects may induce small shifts in the peak positions.
Nevertheless, the qualitative features of the spectra remain meaningful for validating our PPP/pCCD-based models.

For both investigated systems, the LR-pCCD+S(HF) and LR-pCCD+S(pCCD) methods yield qualitatively similar spectra.
For benzo[a]coronene (\ce{C28H14}), calculations display a dominant absorption peak near 310--320~nm, in good agreement with the experimental absorption band reported in the same region.~\cite{bagley2013} Additional weaker features in the 230--370~nm range qualitatively reproduce the multi-band structure observed experimentally.

For naphtho[8,1,2-abc]coronene (\ce{C30H14}), the computed spectra show multiple absorption bands in the 250--430~nm range, with the strongest features near 310--370~nm, consistent with the experimental 
multi-band structure reported in Ref.~\citenum{bagley2013}.


Overall, the LR-pCCD+S approach with the PPP Hamiltonian successfully reproduces the positions of the primary absorption bands of the investigated PAHs.
Differences in relative intensities and detailed band shapes are expected for a semi-empirical model, as the PPP Hamiltonian does not fully account for all-electron correlation effects, vibronic coupling, or solvent interactions.

\subsection{Conclusion and Outlook}
In this work, we present the first incorporation of pCCD-based methods, implemented in the PyBEST software package, with the PPP model Hamiltonian.
The integration of pCCD-based methods and model Hamiltonians yields better scalability and predictive accuracy for large-scale electronic structure studies with small HOMO-LUMO gaps, offering advantages over conventional electronic structure methods.~\cite{pccd-delaram-rsc-adv-2023, pccd-perspective-jpcl-2023, ea-eom-pccd-jpca-2024}

The pCCD approach optimizes all orbitals variationally, resulting in a well-balanced basis across different molecular sizes.
This contrasts with methods like MRCISD, which require a carefully defined active space and are less adaptable to large or diverse systems.
Due to the localized nature of the pCCD-optimized basis, the standard PPP model parametrization, typically defined for canonical HF orbitals, must be reoptimized to ensure accuracy.
Most importantly, our predicted LR-pCCD+S(HF) UV-Vis spectra reproduce the qualitative features of the experimental spectra for Benzo[a]coronene and Naphtho[8,1,2-abc]coronene.
The experimental UV-Vis spectra are predicted qualitatively well within both molecular orbital bases (delocalized and localized). 

In conclusion, this study establishes pCCD/PPP-based methods as an efficient and accurate framework for computing excited-state properties in large $\pi$-conjugated organic systems, such as extended PAHs.
By overcoming the molecular size limitations of traditional ab initio methods and eliminating the need for active space selection, these approaches hold significant potential for guiding the rational design and development of advanced organic optoelectronic materials and devices, including components for next-generation solar cells and organic light-emitting diodes (OLEDs).

Future work will focus on applying modern machine learning techniques to automatically determine optimal parameters for pCCD-based methods, including the $\kappa$ parameter.
This approach is expected to significantly accelerate parameter optimization and enable the study of more realistic models of organic solar cells, thereby going beyond the current limitations of conventional quantum chemistry methods.
\section{Supplementary Material}
The supplementary information provides additional benchmark results.
Specifically, Table S1 of SI.pdf lists the first vertical excitation energies computed using EOM-pCCD+S and EOM-CCSD with the [-2.6/1/11.13] and [-2.4/1/11.13] parameter sets, utilizing both canonical RHF and pCCD-optimized orbitals. Additionally, the file geometry.dat contains the optimized XYZ structures for all molecules investigated in this study.

\begin{acknowledgments}
Z.~K., S.~A., and P.~T.~acknowledge financial support from the SONATA BIS research grant from the National Science Centre, Poland (Grant No. 2021/42/E/ST4/00302). 
Funded/Co-funded by the European Union (ERC, DRESSED-pCCD, 101077420).
Views and opinions expressed are, however, those of the author(s) only and do not necessarily reflect those of the European Union or the European Research Council. Neither the European Union nor the granting authority can be held responsible for them. 
\end{acknowledgments}

\bibliography{rsc.bib}

\end{document}


\renewcommand{\thefigure}{S\arabic{figure}}
\renewcommand{\thesection}{S\arabic{section}}
\renewcommand{\thetable}{S\arabic{table}}
\renewcommand{\tablename}{{Table } \hspace{-0.28cm}}

\begin{center}
\begin{spacing}{2.0}
{\LARGE\bf Efficient and reliable modeling of large $\pi$-electron systems with the Pariser--Parr--Pople Hamiltonian and pCCD-based methods }
\end{spacing}

\vspace{2cm}
{\large 
{Zahra Karimi,$^a$
 Somayeh Ahmadkhani,$^{a}$
 Katharina Boguslawski,$^{a*}$
 and Paweł Tecmer$^{a*}$}
}\\[4ex]

{$^a$\textit{Institute of Physics, Faculty of Physics, Astronomy and Informatics, 
Nicolaus Copernicus University in Toruń, Grudziądzka 5, 87-100 Toruń, Poland}}\\[4ex]

*Email: k.boguslawski@umk.pl, ptecmer@fizyka.umk.pl

\vspace{5cm}

{\bf \Large Supplementary Information}

\vfil

\end{center}

\newpage

\begin{table}[htbp]
\centering
\footnotesize
\setlength{\tabcolsep}{2pt}
\setlength{\arrayrulewidth}{0.4pt}
\caption{First vertical excitation energies (in eV) of selected PAHs computed using the PPP Hamiltonian with different screening parameterizations exploiting canonical RHF and natural pCCD-optimized orbitals.
The mean error (ME), mean absolute error (MAE), and standard deviation (SD) are given with respect to MRSDCI data.
The PPP parameters are labeled as [$t$/$\kappa$/$U$].
MRSDCI reference values are taken from Ref.~\cite{bhattacharyya-2020}. MRSDCI value for coronene 
taken from Ref.~\cite{aryanpour2014}. Experimental values are taken from Refs.~\cite{abouaf2009, bagley2013} for coronene and its derivatives, respectively. }

\vspace{0.5em}
\label{tab:eom_U11}
\begin{tabular}{l|c|cc|cc|cc|cc|cc|cc|c}
\hline
Molecule & MRSDCI & \multicolumn{8}{c|}{EOM-pCCD+S} & \multicolumn{4}{c|}{EOM-CCSD} & Exp\\
\cline{3-10} \cline{11-14}
&&
\multicolumn{4}{c|}{[-2.6/1/11.13]} &
\multicolumn{4}{c|}{[-2.4/1/11.13]} &
\multicolumn{2}{c|}{[-2.6/1/11.13]} &
\multicolumn{2}{c|}{[-2.4/1/11.13]} &
\\
\cline{3-14}
&&
\multicolumn{2}{c|}{RHF} & \multicolumn{2}{c|}{pCCD} &
\multicolumn{2}{c|}{RHF} & \multicolumn{2}{c|}{pCCD} &
RHF & pCCD &
RHF & pCCD &
\\
&&
EE [eV] & OS &
EE [eV] & OS &
EE [eV] & OS &
EE [eV] & OS &
EE [eV] & EE [eV] & EE [eV] & EE [eV]
\\
\hline
$\mathrm{C}_{22}\mathrm{H}_{12}$ & 3.39 & 3.69 & 0.36 & 3.83 & 0.35 & 3.47 & 0.27 & 3.63 & 0.28 & 3.70 & 3.70 & 3.53 & 3.53 & -- \\
$\mathrm{C}_{28}\mathrm{H}_{14}$ & 3.42 & 3.59 & 0.08 & 3.75 & 0.08 & 3.36 & 0.06 & 3.54 & 0.05 & 3.61 & 3.63 & 3.44 & 3.44 & 3.38 \\
$\mathrm{C}_{30}\mathrm{H}_{14}$ & 2.95 & 3.39 & 0.29 & 3.51 & 0.36 & 3.19 & 0.24 & 3.34 & 0.19 &  3.36 & 3.36 & 3.21 & 3.21 & 3.02 \\
$\mathrm{C}_{32}\mathrm{H}_{16}$ & 2.93 & 3.43 & 0.19 & 3.62 & 0.16 & 3.20 & 0.16 & 3.43 & 0.16 & 3.46 & 3.46 & 3.30 & 3.30 & -- \\
$\mathrm{C}_{36}\mathrm{H}_{18}$ & 2.70 & 3.22 & 0.32 & 3.42 & 0.35 & 3.05 & 0.27 & 3.27 & 0.28 & 3.22 & 3.26 & 3.11 & 3.20 & -- \\
$\mathrm{C}_{24}\mathrm{H}_{12}$ & 4.14 & 5.43 & 3.06 & 4.99 & 0.12 & 5.11 & 0.01 & 5.37 & 40.67 &  4.71 & 4.35 & 4.74 & 4.47 & 4.28 \\ \hline
ME    && +0.3 &  & +0.55& & +0.18 & & +0.36 & & +0.39 & +0.40 & +0.24 & +0.26 & \\
MAE   && 0.39 &  & 0.55 & & 0.20 & & 0.36 & & 0.39 & 0.40 & 0.24 &  0.26 & \\
SD    && 0.15  &  & 0.17 & & 0.16 & & 0.18 & & 0.14 & 0.15  &  0.16 & 0.19 & \\ \hline
\hline
\end{tabular}
\end{table}

\bibliographystyle{rsc}
\bibliography{rsc.bib}